\documentclass[a4paper]{article}

\usepackage{INTERSPEECH2020}
\usepackage{amssymb}
\usepackage{multirow}
\usepackage[normalem]{ulem}
\useunder{\uline}{\ul}{}
\usepackage[table,xcdraw]{xcolor}

\usepackage{todonotes}
\usepackage{hyperref}
\usepackage{caption}
\usepackage{subcaption}
\usepackage{diagbox}

\newcommand{\floor}[1]{\lfloor #1 \rfloor}

\title{Surgical Mask Detection with Convolutional Neural Networks and Data Augmentations on Spectrograms}
\name{Steffen Illium, Robert M\"uller, Andreas Sedlmeier and Claudia Linnhoff-Popien}
\address{Mobile and Distributed Systems Group, LMU Munich}
\email{\{steffen.illium,robert.mueller,andreas.sedlmeier,linnhoff\}@ifi.lmu.de}

\begin{document}

\maketitle
\begin{abstract}
In many fields of research, labeled data-sets are hard to acquire. This is where data augmentation promises to overcome the lack of training data in the context of neural network engineering and classification tasks. The idea here is to reduce model over-fitting to the feature distribution of a small under-descriptive training data-set. We try to evaluate such data augmentation techniques to gather insights in the performance boost they provide for several convolutional neural networks on mel-spectrogram representations of audio data.
We show the impact of data augmentation on the binary classification task of surgical mask detection in samples of human voice~(\textit{ComParE Challenge 2020}). Also we consider four varying architectures to account for augmentation robustness. Results show that most of the baselines given by \textit{ComParE} are outperformed.

\end{abstract}

\section{Introduction}
\label{sec:introduction}

In relation to the wide variation of the human voice as well as its differences in expression through loudness, speed and tone, one can image the vast amount of variation present in this data distribution. 
In comparison, the data-set provided in context of \textit{ComParE Challenge 2020} of about \textit{ 10H} of spoken human voice, is a rather small sample out of this total distribution.
We investigate methods of data augmentation to overcome this disadvantage by enhancing the performance of four neural network architectures.
Our motivation to evaluate convolutional models, is the advantage of available choices in data augmentations (raw audio + image representation) and their demonstrated performance of recent years.
Recent research on audio classification has shown the legitimacy of processing mel spectrograms extracted from raw audio signals.

The structure of this paper is as follows: Related work (\autoref{sec:related_work}) presents automatic classification of audio segments by neural networks and strategies of data augmentation. Then our methods of augmentation strategies and model architectures are described in \ref{sec:methods}. 
We then picture the challenge data-set of \textit{ComParE} in \autoref{sec:dataset} and describe our experiment implementation in \autoref{sec:experiments}.
Results are presented in \autoref{sec:results}, which is followed by a discussion in \autoref{sec:conclusion}.

\section{Related Work}
\label{sec:related_work}

\subsection{Audio Classification}
The work at hand can be placed in the field of audio classification which comprises a wide variety of different tasks like multi-label classification to predict notes in musical recordings \cite{thickstun2016learning}, predicting genre tags for songs \cite{dieleman2014end}, environmental sound classification \cite{piczak2015environmental}, or acoustic scene classification \cite{eghbal2016cp}, in which a single label has to be predicted for an entire audio clip.

In recent years, research tackling such audio classification problems has largely shifted from using manually constructed features, e.g. based on mel-frequency cepstral coefficients (MFCC), to using deep neural networks (DNN) in order to automatically learn task-relevant features.
In such approaches of audio classification (when using DNN), it is possible to differentiate further based on the type of input data that is used: i) learning from spectrogram features or ii) direct end-to-end learning from raw audio, i.e. wave-forms.

Learning from spectrogram features, i.e. representing the audio input using time-frequency representations, it is possible to directly leverage architectures that were initially developed for image data processing, such as convolutional neural networks (CNN).
The performance of such for classifying environmental and urban sound clips using log-scaled mel-spectrograms as input is for example investigated in \cite{piczak2015environmental}. The authors show that a deep convolutional model outperforms classical methods that rely on manually constructed features.
\cite{hershey2016cnn} also compute log-scaled mel-spectrograms before evaluating the performance of different DNN-architectures in a video soundtrack classification task.

By contrast, an investigation into the ability of CNN to learn useful features directly from raw audio signals, for an automatic tagging task is presented in \cite{dieleman2014end}. Results show that while it is possible to learn directly from raw audio, networks using spectrogram-based input outperform those using raw audio as input.
Although \cite{dai2016deep} show that it is possible for end-to-end learning approaches to match the performance of CNNs using log-mel spectrograms, the authors only achieve this by using very deep networks, which consequently require a lot of computational power.

An interesting argument that it is not only the training procedure, but the architecture of CNNs by itself that is an important element in the quest for learning good feature representations is presented in \cite{pons2019randomly}.
By evaluating the performance of untrained, i.e. randomly weighted networks, the authors are able to show that excluding any optimization steps, these networks are able to reach impressive accuracy on different audio classification tasks, outperforming an MFCC baseline.

In the work at hand, we chose to follow the route of first extracting mel-spectrograms and using these as input for CNN model architectures.

\subsection{Audio Data Augmentation}

The idea of data augmentation in the field of audio processing and more specific classification tasks is far from new. 
\cite{ko2015audio} for example introduce a combination of vocal tract length perturbation (VTLP), speed perturbation as well as a shift in tempo. 
Perturbations in speed were found to be the most helpful augmentation approach.

Further techniques such as masking or \textit{SpecAugmentation}~\cite{specaug}), which, at first, seem to introduce random shapes of zeros in mel-spectrograms, mask certain areas of a given maximal size of frequency bands and temporal bins. In combination with time warping, it was shown that is is possible to overcome model over-fitting by this kind of data augmentation.

In \cite{dataaug2019classification} data augmentation is added to enhance music genre classification: Loudness, noise introduction time stretching and pitch shifting are applied before CNN models learn the classification tasks.

Research has shown that data augmentation is a valid strategy in the field of machine learning.
We follow this route by implementing and evaluating some promising augmentation techniques 
to overcome the over-fitting problem of the given dataset of \textit{ComParE 2019 challenge} \cite{compare2020}.

\section{Methods}
\label{sec:methods}

\subsection{Data Augmentation}
\label{sec:data_augmentation}
There are five different augmentation approaches we consider for audio data augmentation to enhance prediction performance of the binary classification task given:
\begin{figure*}[htb]
     \centering
     \begin{subfigure}[b]{0.245\textwidth}
         \centering
         \includegraphics[width=\textwidth]{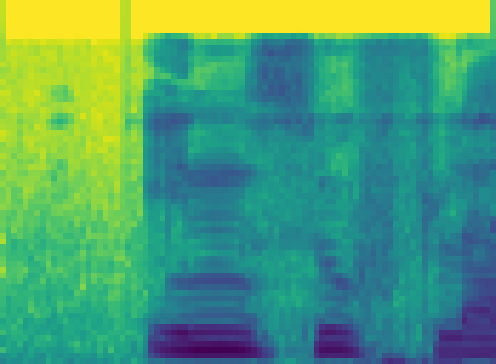}
         \caption{Raw Mel-Spectrogram}
         \label{fig:speed}
     \end{subfigure}
     \hfill
     \begin{subfigure}[b]{0.245\textwidth}
         \centering
         \includegraphics[width=\textwidth]{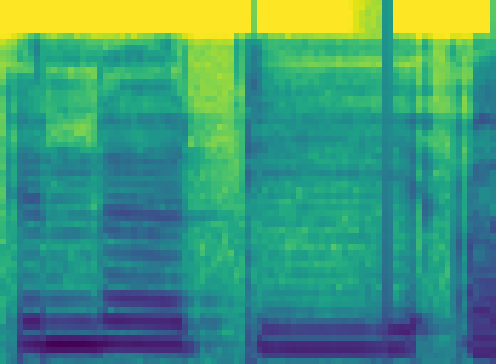}
         \caption{Speed Variation}
         \label{fig:mask}
     \end{subfigure}
     \hfill
     \begin{subfigure}[b]{0.245\textwidth}
         \centering
         \includegraphics[width=\textwidth]{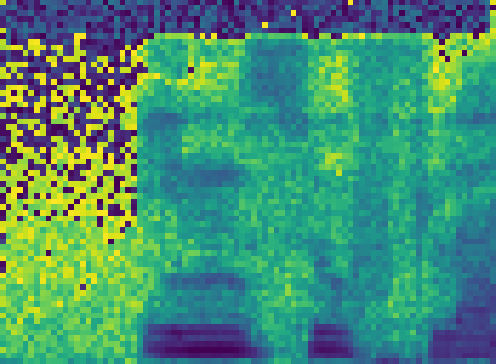}
         \caption{Gaussian Noise Injection}
         \label{fig:noise}
     \end{subfigure}
     \hfill
     \begin{subfigure}[b]{0.245\textwidth}
         \centering
         \includegraphics[width=\textwidth]{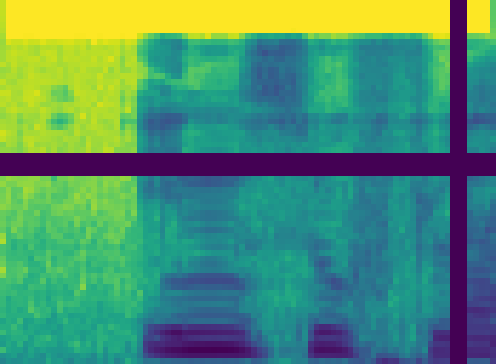}
         \caption{Masking (SpecAugment)}
         \label{fig:spec}
     \end{subfigure}~\\
    \begin{subfigure}[b]{0.245\textwidth}
         \centering
         \includegraphics[width=\textwidth]{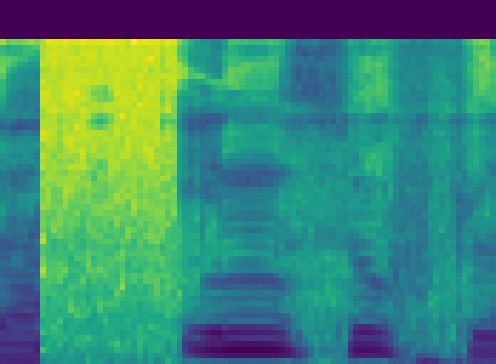}
         \caption{Temporal Shift}
         \label{fig:shift}
     \end{subfigure}
     \begin{subfigure}[b]{0.245\textwidth}
         \centering
         \includegraphics[width=\textwidth]{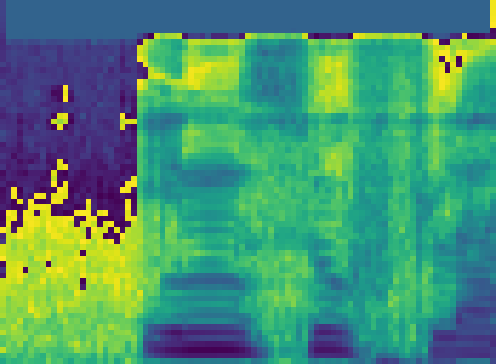}
         \caption{Loudness}
         \label{fig:loudness}
     \end{subfigure}
     \begin{subfigure}[b]{0.245\textwidth}
         \centering
         \includegraphics[width=\textwidth]{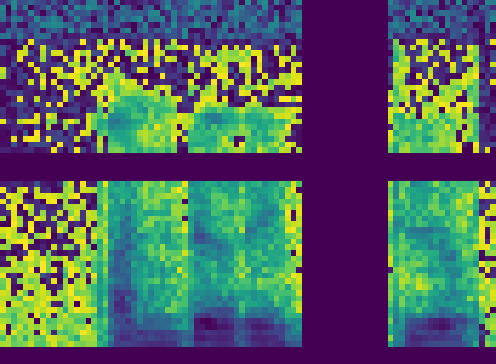}
         \caption{All Combined Augmentations}
         \label{fig:all_combined}
     \end{subfigure}
     \hfill
        \caption{Examples for transformed mel-spectrograms, that are used in model training as given in \autoref{sec:methods}. }
        \label{fig:mel_transformed}
\end{figure*}

\indent\textbf{Speed augmentation}\\
This augmentation processes either speeds up or slows down an audio recording~\cite{ko2015audio}. First we randomly select a starting point $s \sim \mathcal{U}(0, T)$ where T is the number of samples in the recording. Then we sample a window length $w \sim \mathcal{U}(0, 0.4)$. The speed of the samples in the range $[s, \floor{w*T}]$ is subsequently adjusted according to $a \sim \mathcal{U}(0.7, 1.7)$.
While implementations that interpolate data-points in mel-spectrograms exist, we apply our implementation to raw audio data.

\indent\textbf{Loudness Augmentation}
We further diversify the data-set by adjusting the loudness (intensity) of recordings. A loudness factor $l \sim \mathcal{U}(0, 0.4)$ is sampled. This determines how much of the original signal is added back to the original sample ($S + S \times l$, where $S$ is a sample). We want loud signals to get even louder, not just to raise all values of a training sample.

\indent\textbf{Shift Augmentation}\\
To ensure temporal offset, the mel-spectrogram is shifted to either the left or right by $shift \sim \mathcal{U}(0, 0.4)$ percent of its length in the time dimension. 
The direction of the shift is determined by $direction \sim Bernoulli(0.5)$.
Resulting empty data-points $s$ are filled with $zero$. There is also the possibility for Gaussian noise as fill value.

\indent\textbf{Noise Augmentation}\\
Random Gaussian noise, $noise \sim \mathcal{N}(0, 0.4)$, is added ($S+S\times noise$) to the mel-spectrogram as it was shown to improve the robustness of end-to-end trained neural networks models.
Research has shown the benefits not only in audio classification (where recordings can be noise) but also in image processing (were photos can be noisy in low light environments, too).

\indent\textbf{SpecAugment}\\
This procedure describes the \textit{masking} of vertical and horizontal windows with $zero$.
Similar to \cite{specaug}, we first determine a random starting point, $s \sim U(0, T)$, then we determine the size of the window by $w \sim \mathcal{U}(0, 0.2)$. 
All data-points $s in w$ are then filled with $zero$.
This procedure is applied for both axes (time and frequency).
Again, there is the possibility for Gaussian noise as fill value.

We analyze these five data augmentation methods as presented in \autoref{fig:all_combined}, individually and in combination.

\subsection{Model Architectures}
\label{sec:model_architectures}
For our experiments we determined five promising convolution based deep neural network models as classifiers.
For the sake of comparison, we choose the same default parameter settings for all the models, if not specified otherwise. Please note the varying size and introduction of additional parameters through changes in network architecture.
\begin{figure*}[htb]
     \centering
     \begin{subfigure}[b]{0.3\textwidth}
         \centering
         \includegraphics[width=\textwidth]{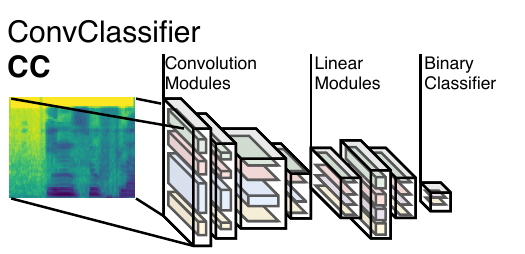}
         \caption{ConvClassifier \textbf{CC}}
         \label{fig:cc_model}
     \end{subfigure}
     \hfill
     \begin{subfigure}[b]{0.3\textwidth}
         \centering
         \includegraphics[width=\textwidth]{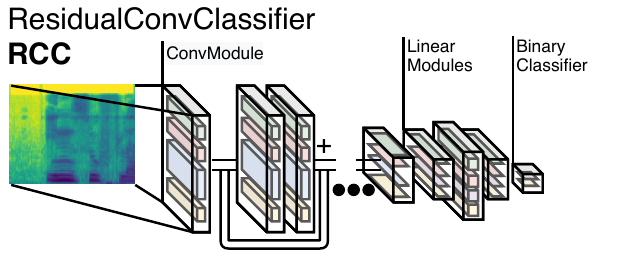}
         \caption{Residual ConvClassifier \textbf{RCC}}
         \label{fig:rcc_model}
     \end{subfigure}
     \hfill
     \begin{subfigure}[b]{0.3\textwidth}
         \centering
         \includegraphics[width=\textwidth]{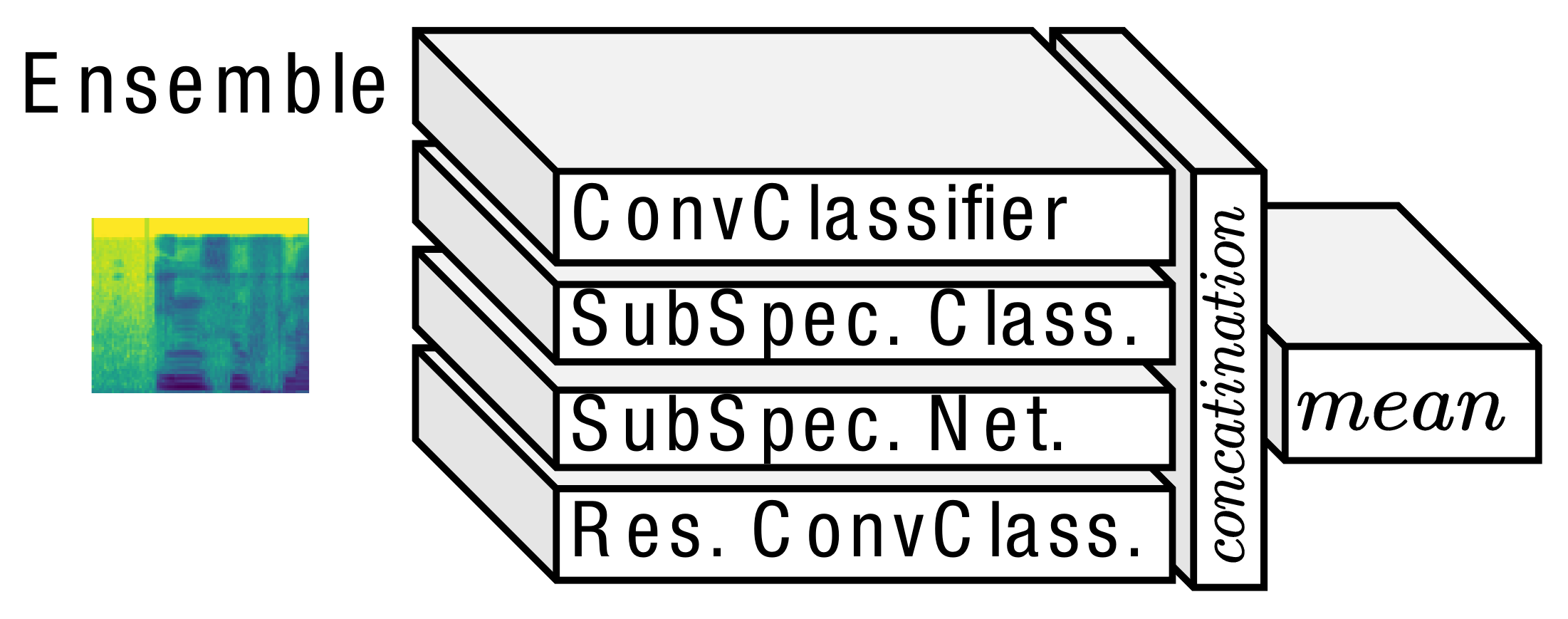}
         \caption{Ensemble}
         \label{fig:e_model}
     \end{subfigure}
    ~\\
    \begin{subfigure}[b]{0.3\textwidth}
         \centering
         \includegraphics[width=\textwidth]{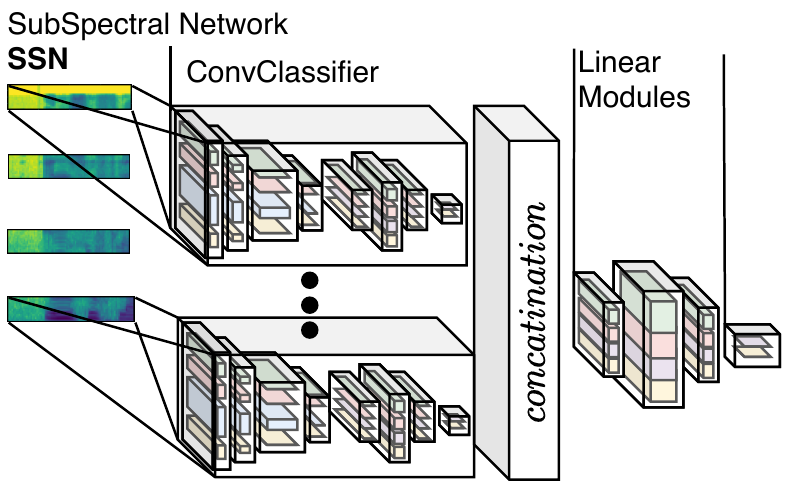}
         \caption{SubSpectral Network \textbf{SSN}}
         \label{fig:ssn_model}
    \end{subfigure}
    \hfill
    \begin{subfigure}[b]{0.3\textwidth}
         \centering
         \includegraphics[width=\textwidth]{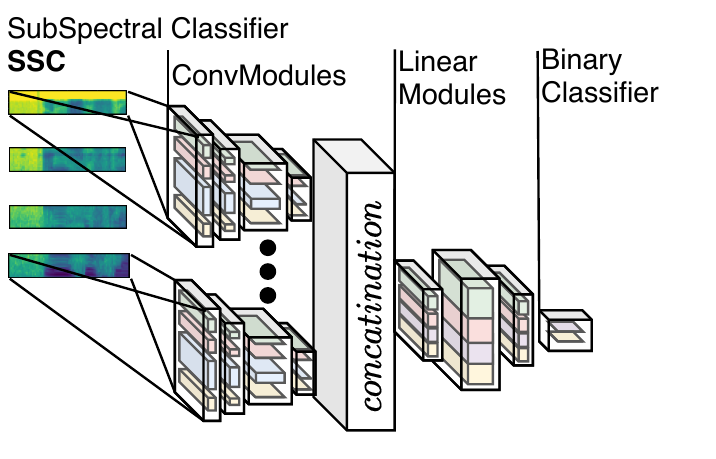}
         \caption{SubSpectral Classifier \textbf{SSC}}
         \label{fig:ssc_model}
     \end{subfigure}
     \hfill
     \begin{subfigure}[b]{0.3\textwidth}
         \centering
         \includegraphics[width=\textwidth]{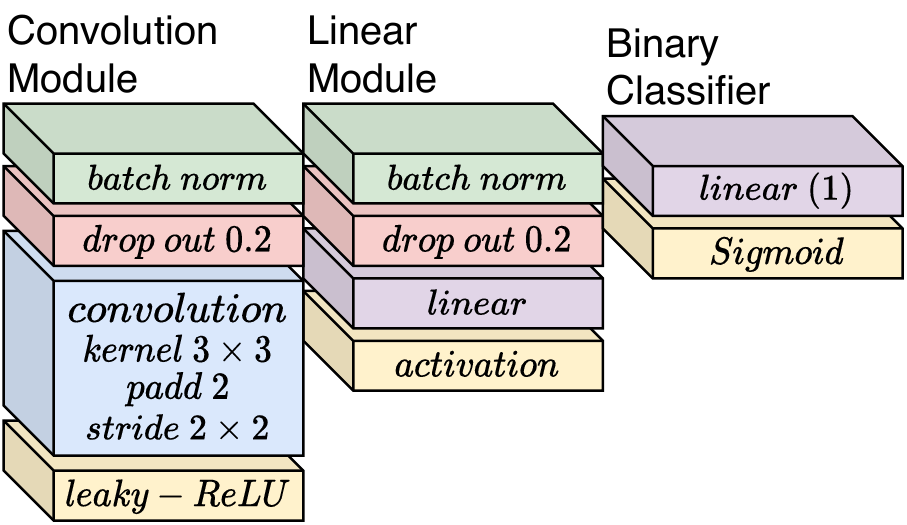}
         \caption{Module Block Description}
         \label{fig:model_blocks}
     \end{subfigure}
     \hfill
        \caption{Model architectures as introduced and described in \autoref{sec:methods}. If not stated otherwise, model  }
        \label{fig:models}
\end{figure*}

\textbf{DefaultNetworkConfiguration}\\
For our experiments we compose four $3\times3$ (\textit{kernel\_size}) subsequent convolution operations with [32, 64, 128, 64] \textit{filters}, respectively.
A striding of two is applied instead of the commonly used max-pooling. 
Additional \textit{zero-padding} of $2\times2$ helps to keep the last tensors' shape at sufficient size. 
The last convolution is followed by four linear layers of different sizes [128, 256, 128, 1], respectively. 
There is a \textit{dropout rate} of $0.2$ before all operations with trainable parameters (convolutions, fully connected linear layer).
Every layer is \textit{leaky-ReLU}~\cite{maas2013rectifier} activated while tensors are \textit{batch-normalized}~\cite{ioffe2015batch}.
The last single neuron layer represents an exception as it is \textit{Sigmoid} activated for the binary classification task.
The classifier output is then evaluated against given training labels by a \textit{BinaryCrossEntropy-Loss} (BCE).

\textbf{ConvClassifier (CC)}\\
This rather classic model \autoref{fig:cc_model} combines state of the art strategies in image classification in the field of machine learning. Four convolution stages reduce the image's spatial resolution while learning to activate feature maps that are growing in size. 
Additional fully connected layers (or linear) learn the main features within the training samples. 
This models follows the architecture that was proposed as our \textit{default network configuration}.

\textbf{SubSpectralNet (SSN)}\\
In reference to~\cite{phaye2019subspectralnet}~we train a combination of four different models based on our \textit{default architecture}. 
Each model is applied to a different non-overlapping number of mel-bands $=64/4$, depicted in \autoref{fig:ssn_model}.
Those concatenated band-wise predictions (Sigmoid activated prediction, not the logit) are then processed by a global classifier sub-network (four fully-connected layers with [128, 256, 128, 1] neurons, respectively).
Further parameters are the same as in our \textit{default configuration}, this includes activation, batch-normalization and position and rate of dropout.
All Sigmoid classifier outputs (band-wise predictions + sub-network prediction)  are individually evaluated, each by a BCE-loss. 
Gradients for model training are finally calculated based on the mean of these losses.

\textbf{SubSpectralClassifier (SSC)}\\
In addition to the SSN model architecture we define another sub-spectral network.
This time the global classifier sub-network has to learn which features to use.
The default convolution stack is applied in a band-wise fashion (like before) while the global classification sub-network processes the concatenated output of all last convolution operation, pictured in~ \autoref{fig:ssc_model}. 
Neurons for the fully-connected classifier are as before [128,256,128,1] while the very last layer is Sigmoid activated.
BCE-Loss is used for classifier output evaluation.

\textbf{ResidualConvClassifier (RCC)}\\
Similar to~\cite{ford2019deep, ren2019spec} we combine residual skip connections around blocks of convolutional operations (\autoref{fig:rcc_model}). 
For this, we modify the \textit{default architecture} by introducing a residual block of two similar shaped convolution operations in between every single convolution operations respectively. 
The fully-connected classifier follows our \textit{default architecture}.

\begin{table*}[htb]
\centering
\caption{Result comparison: We present model performance under the influence of a variety of data augmentation strategies.}
\label{tab:main_result}
\begin{tabular}{l|cccc|l}
         & \begin{tabular}[c]{@{}l@{}}ConvClassifier\\ (CC)\end{tabular} & \begin{tabular}[c]{@{}l@{}}SubSpectral\\ Network (SSN)\end{tabular} & \begin{tabular}[c]{@{}l@{}}SubSpectral\\ Classifier (SSC)\end{tabular} & \begin{tabular}[c]{@{}l@{}}Residual Conv-\\ Classifier (RCC)\end{tabular} &\textbf{ Max.} \\ \hline
         
Raw      & $64.71\pm0.30$       & $63.54\pm0.26$    &$65.82\pm0.39$         &$64.53\pm0.53$         &$65.82$          \\ \hline

Speed    &$64.71 \pm 0.40$      &$63.54\pm0.91$     &$66.40\pm0.34$          &$64.07\pm0.77$         &$66.40$          \\

Noise    &$ 63.62\pm 0.22 $     &$61.35\pm0.67$     &$64.48\pm0.46$         &$63.38\pm0.41$         &$64.48$          \\

Loudness &$ 64.31 \pm 0.43$     &$62.87\pm0.64$     &$65.65\pm0.50$          &$64.14\pm0.37$         &$65.65$          \\

Shift    &$ 65.40 \pm 0.55$     &$66.31\pm0.70$      &$\textbf{68.20}\pm0.44$          &$64.38\pm0.78$         &$\textbf{68.20}$          \\

Masking  &$ 64.27 \pm 0.34 $    &$62.27\pm0.33$     &$65.10\pm0.32$         &$64.30\pm0.45$          &$65.10$          \\ \hline

Combined &$ 65.03\pm0.41$       &$63.49\pm0.79$     &$66.35\pm0.34$         &$64.12\pm0.42$         &$66.35$    \\\hline

\textbf{Max.}     &$65.40$               &$66.31$            &$\textbf{68.20}$       &$64.38$ \\ \hline

\begin{tabular}[c]{@{}l@{}}Model \\ Parameters \end{tabular} & 633,219 & 1,801,621 & 1,533,321 & 1,095,171
\end{tabular}
\end{table*}

\section{Dataset}
\label{sec:dataset}

We focused on the augmentation of audio recordings, included in the Mask Augsburg Speech Corpus (MASC) as given as part o\textit{f ComParE 2020 Challenge}~\cite{compare2020}. 
The original dataset consists of 10 h 9 min 14 sec of audio recordings.
There are a total of 32 speakers (German native, 16 f, 16 m, age from 20 to 41 years, mean age 25.6 years $\pm 4.5$), which are performing different tasks with and without wearing a surgical mask. 
It is stated that all audio samples were recorded in a sound-proof audio studio with proper equipment at 44kHz which have been down-sampled and converted to 16 kHz and mono/16 bit. 
The recordings were pre-segmented into chunks of 1 sec duration without overlap.
As motivated in \autoref{sec:introduction} and proposed in \autoref{sec:methods} we transform the raw audio data regarding speed of various intensity and length.
Hence, the amount of training data samples quadrupled, whereas the validation and test data-sets both stay untouched. 
Please note that at least one raw (not transformed) sample of the initial training data is always maintained.

\section{Experiments}
\label{sec:experiments}
Our experiments are conducted under controlled settings. 
Models have been trained on the same training data as we used only fixed seed random operations (along python, numpy, pytorch).
Data augmentation is 
performed as proposed in \autoref{sec:methods}.
We are especially inspired by the implementation of \textit{nlpaug}~\cite{ma2019nlpaug}.
Parameters for data augmentation are chosen through the experience of various experiments, as follows:  $\text{speed factor} = 0.7$, $\text{speed ratio} = 0.3$, $\text{masking ratio} = 0.2$, $\text{noise ratio} = 0.4$, $\text{shift ratio} = 0.3$, $\text{loudness ratio} = 0.4$. 
They stay the same throughout all combination of models and seeds in training (no transformation is applied in validation).
Mel-spectrograms are extracted at a window $\text{hop-length=256}$ at $\text{n-fft=512}$ transformations for $\text{n-mels=64}$ bands.
This results in a sample shape of $64 \times 87$.
All samples are then locally normalized (zero mean, unit variance).

Data processing and augmentation is performed as follows: 
First, we transform the train dataset by applying the \textit{speed augmentation} as it needs to be applied on the raw audio source.
We then extract log mel-spectrogram from the original audio sample (by \textit{librosa}\footnote{https://librosa.github.io/ - 10.5281/zenodo.3606573}) for training as well as for validation data, which is computationally expensive.
For this reason, computation of spectrograms is performed once per training (seed), rather than per epoch. 
Spectrograms are then transformed into log scale, inverted (dark=low energy) and stored as single channel 8-bit grey-scale image (values between 0-255).
All further augmentations are applied in real-time, which is done before local normalization takes place.
The dataset is randomly sampled and batched for the models.

Training procedure is applied as follows: We trained each model on five different seeds by back-propagation through \textit{Adam} optimizer \cite{kingma2014adam} at a learning-rate of $lr= 1e-4$, weight-decay $= $ for $51~\text{epochs}$ ($146$ batches per epoch, batch-size of $200$).
Losses are calculated as given in \autoref{sec:methods} by BCE as a binary classification task.
All scores are then measured by the \textit{unweighted average recall}~(UAR)~(cf.~\autoref{eq:uar_ber}
as required by the rules of \textit{ComParE 2020 Challenge}~\cite{compare2020}.
It is also known as the \textit{Balanced Error Rate}~(BER)~\cite{rosenberg2012classifying}.
\begin{equation}
    UAR = 0.5 \times \bigg(\frac{TP}{FN+TP} + \frac{TN}{TN+FP}\bigg)
    \label{eq:uar_ber}
\end{equation}

\section{Results}
\label{sec:results}
\begin{table}[ht]
\centering
\caption{Comparison between different models. Performance is measured in terms of the UAR.}
\label{tab:results}
\begingroup
\setlength{\tabcolsep}{6.25pt} 
\renewcommand{\arraystretch}{1.25} 
\begin{tabular}{l|cc}
\hline
\multicolumn{1}{c|}{\multirow{2}{*}{\textbf{Baseline Models}}} & \multicolumn{2}{c}{\textbf{UAR}}                                              \\
\multicolumn{1}{c|}{}                                          & \multicolumn{1}{l}{\textbf{Dev}} & \multicolumn{1}{l}{\textbf{Test}} \\ \hline
DeepSpectrum + SVM (ResNet50)                                  & 63.4       & 70.8  \\
S2SAE + SVM (Fused)                                            & 64.4       & 66.6  \\
ComParE functionals + SVM (C=$10^{-3}$)                        & 62.3       & 67.8  \\
ComParE BoAW + SVM (N=$2$k)                                    & 64.2       & 67.7  \\
Fusion                                                         & -          & 71.8  \\
\cline{1-1}
\multicolumn{1}{c|}{\textbf{Our Models}}                       & \multicolumn{1}{l}{}            & \multicolumn{1}{l}{}\\ 
\cline{1-1}
ConvClassifier + Shift                                         & 65.4      & -       \\
SpectralNet Network + Shift                                    & 66.3      & -       \\
SpectralNet Classifier + Shift                                 & 68.2      & 71.5    \\
Residual Conv. Classifier + Shift                              & 64.5      & -       \\
Ensemble over best Models; Mean Vote                           & 68.0      & -       \\ 
\hline
\end{tabular}
\endgroup
\end{table}
In \autoref{tab:main_result} we present the influence of different augmentation techniques as well as their combinations in contrast to model training on raw, non-augmented data-sets.
We also look at different model results for those augmentation methods, at varying trainable parameter sizes. 
The maximum reached \textit{UAR-scores} (cf.~\autoref{sec:experiments}) over all model instances as well as its variance are reported to a total of 100.
We found it surprising to see that a conventional CNN ($3\times3$ kernel applied non-overlapping $stride = 2$) on normalized mel-spectrograms already reached a score which could compete with the baseline UAR score (devel.).
Data augmentation just minimally enhanced these results by about 1\%.
Temporal Shift was found to be the overall best and most reliable augmentation strategy, compared to the other domain specific augmentation methods implemented.
The proposed SubSpectral Classifier (SSC) (inspired by SSN~\cite{phaye2019subspectralnet}) achieved promising results on raw samples, which could be further amplified through augmentation by temporal shift.
All models are robust to the changes we applied to training data and stayed within their range of performance (varying up to 5\%).
The wild combination of available augmentations did not prevail, which is rather not surprising. 
The Residual ConvClassifier (RCC) seems to be more stable as a model itself as data augmentation did not influence the model performance noticeably.

\section{Conclusion}
\label{sec:conclusion}
In this work, we demonstrated the influence of data-augmentation on the binary classification task of surgical-mask detection in context of the \textit{ComParE Challenge 2020}~\cite{compare2020}.
We not only implemented and evaluated different augmentation methods, but also showed their influence on different model architectures.
We found the combination of temporal shift with standard CNN architectures to be a competitive strategy.
Further, our proposed SubSpectral Classifier (SSC) achieved better results, while performing at a similar variance in evaluation.
The implemented residual network (RCC) was the overall most robust architecture, as any influence of data-augmentation was not measurable. Not even when combining all available perturbations.

In the future, we aim to evaluate a combination of models, especially with SubSpectral variants. Model ensembles (depicted in Figure \ref{fig:e_model}) showed promising results compared to the given baselines (cf. Table \ref{tab:results}), but were not able to prevail against the SSC model in mean or majority voting.

\newpage
\bibliographystyle{IEEEtran}
\bibliography{mybib}

\end{document}